\documentclass[11pt]{article}
\usepackage{nodalida2017}
\usepackage{mathptmx,amsmath,amssymb,amsthm,accents,textcomp}
\newtheorem{proposition}{Proposition}

\newtheorem{openproblem}{Open Problem}
\usepackage{url}
\usepackage{latexsym}
\usepackage{tikz-dependency}
\usepackage{tikz}
\usetikzlibrary{arrows}

\title{The Power of Constraint Grammars Revisited}

\author{Anssi Yli-Jyr\"a \\ University of Helsinki, Finland  \\ {\tt anssi.yli-jyra@helsinki.fi}
}

\date{}

\begin{document}
\maketitle
\begin{abstract}
  Sequential Constraint Grammar (SCG) \cite{Karlsson-1990} and its
  extensions have lacked clear connections to formal language theory.
  The purpose of this article is to lay a foundation for these
  connections by simplifying the definition of strings processed by
  the grammar and by showing that Nonmonotonic SCG is undecidable and that derivations similar to the
  Generative Phonology exist.  The current investigations propose resource bounds that restrict the generative power of SCG to a subset of context sensitive languages and
  present a strong finite-state condition for grammars as wholes.  We show that a grammar is equivalent to a finite-state
  transducer if it is implemented with a Turing machine that runs in
  $o(n\log n)$ time. This condition opens new finite-state hypotheses
  and avenues for deeper analysis of SCG instances in the way inspired by Finite-State Phonology.  
\end{abstract} 

\section{Introduction}

\newcite{Lindberg1998}, \newcite{Lager-Nivre-2001:parofspetag:inc} and
\newcite{Inari2016} have analyzed the Sequential Constraint Grammar
(SCG) \cite{Karlsson-1990} from the logical point of view, proposing
that the rules can be expressed in first-order Horn clauses,
first-order predicate logic or propositional logic.  However, many
first-order logical formalisms are themselves quite expressive as
Horn-clauses are only semi-decidable and first-order logic is
undecidable, thus at least as powerful as SCG itself.  Propositional
logic is more restricted but does not help us to analyse the
expressive power of SCGs and to prove the finite-stateness of grammars.

Instead of just reducing SCG to undecidable or otherwise powerful
formalisms, we are interested in the ultimate challenge that tries to
prove that a \emph{practical} grammar is actually reducible to a
strictly weaker formalism.  This goal is interesting
because this kind of narrowing reductions have been proven extremely valuable.  For
example, the proof that practical grammars in Generative Phonology are
actually equivalent to finite-state transducers has turned out to be
a game-changing result.  In fact, the reduction gave birth to the
influential field of Finite-State Phonology.

It is noteworthy that prior efforts to analyse SCG in finite-state
terms have focused on the finite-state nature of individual and
parallel rules \cite{Peltonen-2011,Hulden-2011,YJ2011}.  The efforts have mostly ignored the generative power of the grammar system as a whole and that of practical grammar instances.  

In this paper, we are aiming to Finite-State Syntax through
reductions of practical SCGs.  To set the formal framework, we have to
start, however, from the total opposite: we show first that the
simplified formalism for Nonmonotonic SCGs is Turing equivalent and
thus similar to Generative Phonology
\cite{SPE,Ristad:1990:CSG:981823.981853} and Transformational Grammar
\cite{Chomsky-1965:aspofthethe:boo,Peters-Ritchie-1973}.  This
foundational result gives access to the large body of literature of
bounded Turing machines and especially to Hennie machines that run in
$O(n)$ time and are equivalent o finite-state machines.  Then the Gap
Theorem \cite{Trakhtenbrot} gives us access to a looser bound $o(n
\log n)$ whose reasonable approximations are sufficient and decidable
conditions for finite-state equivalence.  We present some ways in
which these bounds can be related to SCG parsing.

The article is structured as follows.  Section 2 describes the
alphabets, the strings and the derivation steps in SCG parsing.  In
Section 3, these are used to show Turing equivalence of SCGs.  In next
two sections, simple bounds are introduced and elaborated further to
obtain specific conditions for finite-state equivalence of grammars.
Further links to formal language theory and two important open
problems are presented in Section 6.  Then the paper is concluded.

\section{SCG as a "Phonological" Grammar}

In the SCG literature, morphosyntactic readings of tokens are usually
represented as tag strings like \texttt{"<went>" "go" V PAST}.  The
tag strings are now viewed as a compressed representation for a huge
binary vector $(f_0,f_1,f_2,...,f_k,...)$.  The semantics of the
grammar ignores some tags and considers only $k$ tags declared in
advance in the grammar.  These $k$ tags or features distinguish
readings from each other and define the \emph{reading alphabet}
$\Sigma=2^k$.

An ambiguous token has more than one reading associated to it.  The
elements of the \emph{cohort alphabet} $\mathcal{P}(\Sigma)$ are
called cohorts.  This alphabet is the powerset of the reading
alphabet.  Only a small subset of all possible cohorts occur in
practice.

The input of an SCG is produced by a deterministic
finite-state function, $Lexicon^*: T^* \rightarrow
(\mathcal{P}(\Sigma))^*$, that maps token strings to \emph{lexical
  cohort strings} of the same length.  This function is the concatenation
closure of the function $Lexicon: T \rightarrow (\mathcal{P}(\Sigma))$
that maps every token to a cohort.

Since the image of each token is a set of strings, $Lexicon$ is
internally a nondeterministic lexical transducer
\cite{Karttunen-1994:conlextra:inp,Chanod-Tapanainen-1995:alexintfor:tec}, but the image of each token is viewed externally as a \emph{symbol} in $\mathcal{P}(\Sigma)$, making $Lexicon$ a one-valued function.

An SCG processes the lexical cohort string by iterated application of
a derivation step $\Rightarrow:(\mathcal{P}(\Sigma))^* \rightarrow
(\mathcal{P}(\Sigma))^*$ that affects one cohort at a time.
The contexts conditions of each derivation step are normally defined
using an existing SCG formalism for contextual tests.  Monadic Second
Order Logic \cite{Buchi-1960:weasecariand:art,Elgot-1961:decprooffin:art,Trakhtenbrot-1961}
provides an alternative formalism that can express all finite state
languages over $\mathcal{P}(\Sigma)$.

\begin{figure}\centering%
  \scalebox{.5}{\includegraphics{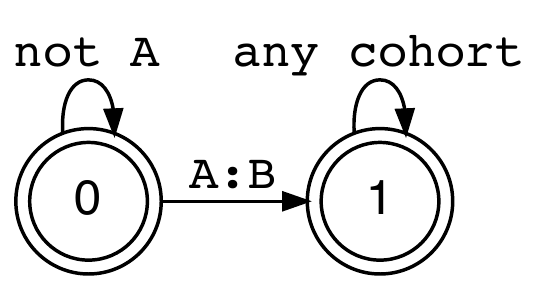}}
  \caption{A simple $\Rightarrow$ relation as an FST\label{deriv}}
\end{figure}

The parser defines the parsing strategy that resolves the conflicts
between rules that could be applied simultaneously.  A typical
strategy chooses always the most reliable rule and the leftmost target
position.  When the plain contextual tests are combined with the
application strategy, we obtain a total functional transducer
\cite{Skut:2004:BCR:1220355.1220384,Yli-Jyra-2008glc,Hulden-2009}.  E.g., the transducer in
Fig. \ref{deriv} is total and replaces A by B in the first possible
occurrence position.

The semantics of an SCG grammar $G$ is defined as the
relation
\[
[[G]] = \{ (i,o) \mid i \in \mathcal{P}(\Sigma)^*, o \in I, i \Rightarrow^* o \}
\]
where $I\subseteq \mathcal{P}(\Sigma)^*$ as $\{x\mid (x,x) \in
\Rightarrow\}$.  This semantics makes SCG grammars similar to grammars
in Generative Phonology \cite{SPE} as both grammars relate the lexical string into some kind of output string by applying a sequence of alternation rules.

\section{Nonmonotonic SCG}

Two recent SCG implementations \cite{Tapanainen-1996,Didriksen-2017}
are \emph{nonmonotonic}: they do not always reduce the input but they
can insert tags, readings and even cohorts.  In this section, we study
the expressive power of such SCGs.

\subsection{Minimal Definition}

For the sake of minimality, we define the Nonmonotonic SCG (NM-SCG) as
a rule system that supports the following kinds of local
transformation rules:
\begin{itemize}
\item \texttt{REPLACE (\textit{old}) (\textit{new})
  (\textit{cond})$^+$}\\[-5ex]
\item \texttt{INSCOHORT (\textit{targ}) (\textit{cond})$^+$}\\[-5ex]
\item \texttt{REMCOHORT (\textit{targ}) (\textit{cond})$^+$}
\end{itemize}
The first rule template in the above replaces the leftmost cohort
containing the reading \texttt{\textit{old}} with a cohort that
contains the reading \texttt{\textit{new}} if the relative context
condition \texttt{\textit{cond}} is satisfied.  The familiar
\texttt{SELECT} and \texttt{DELETE} rules are seen as shorthands for
sets of \texttt{REPLACE} rules.  The second and the third rule
templates are used to insert or remove a target cohort matching the
pattern \texttt{\textit{targ}} when the condition
\texttt{\textit{cond}} is satisfied.  The plus ($^+$) indicates that
more than one condition can be present.

Our simplified context conditions are of the form $\texttt{($d$
  \textit{tags})}$ or $\texttt{($d$ NOT \textit{tags})}$ where the
first tests the presence of the pattern \texttt{\textit{tags}} in the
relative cohort location $d$.  The second is true when the location does not contain the pattern.

\subsection{One-Tape Turing Machine}

\begin{figure}[h!]
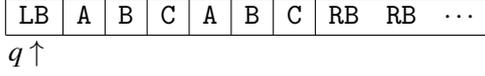

\[
\begin{array}{|c|c|c|c|c|c|c|ccc}
  \hline
  \texttt{LB} & 
  \texttt{A} & 
  \texttt{B} & 
  \texttt{C} & 
  \texttt{A} & 
  \texttt{B} & 
  \texttt{C} & 
  \texttt{RB} &
  \texttt{RB} &
  \cdots\\
  \hline
  \multicolumn{1}{c}{\!\!\!\!\!q\uparrow} \\[-3ex]
\end{array}
\]
\caption{A one-tape Turing machine\label{tm}}
\end{figure}

A one-tape deterministic \emph{Turing machine} (TM) has a finite
control unit and an infinite rewritable tape with a pointer
(Fig. \ref{tm}).  A configuration of the machine consists of the
current state $q$, the current pointer value and the contents of the
working tape.

The tape is divided into squares that hold a left boundary
\texttt{LB}, a right boundary \texttt{RB}, or a symbol from the
tape alphabet $\Omega$.  Given the input string $x\in \Omega^*$, the
first square of the tape is pointed and the tape is initialized 
with the prefix $\texttt{LB}x\texttt{RB}$ that is followed by an
infinite number of right boundary symbols.  

The control unit is a deterministic finite automaton where each
transition $s \overset{(A,B,d)}\rightarrow t$ specifies the source
state $s$, the target state $t$, the input symbol $A$, the output
symbol $B$, and a head move $d\in \{-1,0,1\}$.  On each transition,
the machine overwrites the symbol $A$ in the pointed square with the
symbol $B$, changes its state from $s$ to $t$ and then moves the
pointer $d$ steps to the right.  All but the leftmost
square are over-writable ($B=A$ if $A=\texttt{LB}$), but the machine never moves beyond the first right boundary without overwriting it with a tape symbol and never writes \texttt{RB} between two tape symbols.

The computation of the machine starts from state $q_0$.  At each step,
the machine takes the next transition based on the current state and
the currently pointed symbol on the memory tape.  The computation
continues as long as the next transition is defined and then halts by
reaching a state from which there is no transition on the current
input.  If the halting state is among the final states $F$, the
machine accepts the input contents and relates it with the string
$x'\in \Omega$ stored to the memory tape.  Otherwise, the machine
either gets stuck to an infinite computation or gives up, leaving
some ill-formed string to the memory tape.

\subsection{Reduction to Nonmonotonic SCG}

Now we show that any one-tape Turing machine can be simulated with a
nonmonotonic SCG.

In our simulation, each square in the initial portion of the memory
tape corresponds to a cohort in the input.  Each cohort is a singleton
set in $\mathcal{P}(\Sigma)$ i.e. represents just one reading in
$\Sigma$.  Each reading is a collection of positive features from
$\Phi$.  These features include the tape symbols $\Omega$, the
boundary symbols $\{\texttt{LB}$, $\texttt{RB}\}$, and the markers
that that we need to keep track of the computation steps.

The pointed square corresponds to a cohort that contains a marker.
Since SCG can change only one cohort at a time, movement of the
pointer involves two temporarily marked positions and markers: the
first indicates the previously pointed square and the second indicates
the new pointed square.  One marker represent the source state and the
other represents the transition in progress.

A transition $q\overset{A,B,d}\rightarrow r$, $\texttt{RB}\notin
\{A,B\}$, corresponds to a sequence of three rule applications that
change one cohort at a time.  Since the set of transitions, the sets
of states $Q$ and the tape alphabet $\Omega$ are finite, each step is
described with a finite set of non-monotonic SCG rules:
\begin{enumerate}
  
\item Given the state marker $\texttt{Q}_q \in \{\texttt{Q}_s \mid
  s\in Q\}\subseteq \Phi$ in cohort $i$ and no other marked cohorts,
  add a transition marker $\texttt{T-$q$-$A$}\in \Phi$ to cohort
  $i+d$ that previously contains a tape symbol $C\in \Omega$:
  \[
  \texttt{\small REPLACE
    ($C$) ($\texttt{T-$q$-$A$}$ $C$) ($-d$ Q$_q$ $A$)}
  \]
\item Given a transition marker $\texttt{T-$q$-$A$}$ in cohort $i+d$,
  overwrite, in cohort $i$, the reading containing the tape symbol $A$
  and the state marker $\texttt{Q}_q$ with a reading containing the
  tape symbol $B$:
  \[\texttt{\small REPLACE (Q$_q$ $A$)
    ($B$) ($d$ $\texttt{T-$q$-$A$}$)}
  \]
\item When no state marker is present, replace the transition marker
  $\texttt{T-$q$-$A$}$ with the marker for the target state
  $\texttt{Q}_r$ while keeping the remainder $C\in\Sigma$ in the
  changed cohort:
  \[\texttt{\small REPLACE ($\texttt{T-$q$-$A$}$ $C$) (Q$_r$ $C$) ($-d$ NOT Q$_q$)}
  \]
\end{enumerate}
A transition $q\overset{\texttt{RB},A,0}\rightarrow r$, $A\in \Omega$,
corresponds to the application of rules:
\begin{gather*}
  \texttt{\small ADDCOHORT ($\texttt{T-$q$-RB}$ $A$) (1 Q$_q$ RB)}\\[-1ex]
  \texttt{\small REPLACE (Q$_q$ RB) (RB) (-1 $\texttt{T-$q$-RB}$ $A$)}\\[-1ex]
  \texttt{\small REPLACE ($\texttt{T-$q$-RB}$ $A$) (Q$_r$ $A$) (1 NOT Q$_q$ RB)}
\end{gather*}
When the previous cohort contains tape symbol $C\in \Omega$, a
transition $q\overset{A,\texttt{RB},-1}\rightarrow r$, where $A\in
\Omega$, corresponds to the application of rules:
\begin{gather*}
  \texttt{\small REPLACE ($C$) ($\texttt{T-$q$-$A$}$ $C$) (1 Q$_q$ $A$) (2 RB)}
\\[-1ex]
  \texttt{\small REMCOHORT ($\texttt{Q}_q$ $A$) (1 $\texttt{T-$q$-$A$}$)}
  \\[-1ex]
  \texttt{\small REPLACE ($\texttt{T-$q$-$A$}$ $C$) (Q$_r$ $C$) (1 NOT Q$_q$ $A$)}
\end{gather*}
Transitions $q\overset{\texttt{RB},A,-1}\rightarrow r$,
$q\overset{\texttt{RB},A,1}\rightarrow r$ and
$q\overset{A,\texttt{RB},0}\rightarrow r$, $A\in \Omega$, reduce 
to a sequence of two transitions.

The SCG parser halts when the tape contents does not trigger any of
these rules that simulate transitions.  The simulation accepts the
input if some cohort contains a marker $\texttt{Q}_q$ such that $q\in
F$.  

\begin{proposition}
  NM-SCGs can simulate TMs.
\end{proposition}
Since NM-SCG is itself an algorithm, we have:
\begin{proposition}
  There is a one-tape deterministic TM that implements the
  NM-SCG parser.
\end{proposition}
\begin{proposition}
  NM-SCGs are equivalent to TMs.
\end{proposition}

\section{Bounded Nonmonotonic SCGs}

The undecidability of Nonmonotonic SCG creates a need to restrict the
formalism in ways that ensure decidability.  In this section, we
propose two parameters that set important bounds on the resources available to grammars.

\subsection{The $O(n)$ Space Bound}

The \emph{fertility} $f\in \mathbb{N} \cup \{\infty\}$ of a
nonmonotonic SCG grammar is the maximum number of new cohorts that
each the grammar inserts before any of the $n$ cohorts in the original
sentence (with \texttt{RB}).  Note that fertility $f > 0$ implies nonmonotonicity.

\begin{proposition}
  In finite-fertility SCGs, the length $\ell$ of the output string is
  linearly bounded.
\end{proposition}

The bounded length of the cohort string is an important restriction to
Nonmonotonic SCGs because it ensures that any infinite loop in the
computation can be detected after a bounded number of computation
steps because the number of distinct tape contents is bounded.

\begin{proposition}
  The termination of a finite-fertility SCG is decidable.
\end{proposition}

We also know that the
preconditions of each rule can tested with a finite automaton and that
the actual effect on the target cohort is a functional finite-state
computation that can be implemented in linear space according to the
length of the cohort string.

\begin{proposition}
  The space requirement of a finite-fertility SCG is linear to the
  maximum length of the cohort string during the derivation.
\end{proposition}

A deterministic linear-bounded automaton (DLBA) \cite{LBA} is a
special case of Turing machines with the restriction that the right
boundary is fixed and cannot be overwritten.  The LBA computations can
be initialized so that the space available for storing the cohort
string is linearly bounded by the length of the initial cohort string.

\begin{proposition}
  A nonmonotonic SCG with finite fertility is simulated by an DLBA.
\end{proposition}

The power of DLBAs is restricted to a strict subset of context-sensitive
languages \cite{Kuroda-1964}.

\begin{proposition}
  The cohort language accepted by a finite-fertility SCG is context
  sensitive.
\end{proposition}

\subsection{The $O(n^2)$ Time Bound}

By studying only monotonic SCGs with the reading count $r$ in cohorts,
and the sentence length $n$ (including \texttt{RB}), \newcite{Tapanainen-1999:parintwofra:phd}
has given a lower bound for the parsing time:
\begin{proposition}[Tapanainen 1999]
  Any monotonic SCG performs $O(nr)$ rule applications.
\end{proposition}

The \emph{volume} $v \in \{1,2,...\} \cup \{\infty\}$ of cohorts is
a parameter that tells the maximum number of operations that can be
applied to any cohort.  This new notion is a nonmonotonic
generalization of the maximum number of readings in one cohort.
Finite volume basically turns every finite fertility SCG into a
monotonic SCG.

Finite fertility helps us to generalize the above proposition to
nonmonotonic SCGs.

\begin{proposition}
  Any NM-SCG performs $O((1+f)nv)$ rule applications.
\end{proposition}

Assuming again that any rule of the grammar can be applied in linear
time according to the number of cohorts, we obtain a time complexity
result:

\begin{proposition}
  Any NM-SCG runs in $O((1+f)^2n^2v)$ time.
\end{proposition}

\section{Finite-State Hypotheses}

A deterministic linear bounded automaton is a special case of one-tape
deterministic Turing machines that gives us a context where many
interesting conditions for finite-stateness aka regularity become applicable.

\subsection{The $o(n\log n)$ Time Bound}

\newcite{Hennie-1965} showed that a deterministic one-tape TM running
in $O(n)$ is equivalent to a finite automaton.  By defining the
relation between the initial and final tape contents, we can extend
Hennie's result to regular relations:
\begin{proposition}[Hennie 1965]
  A one-tape deterministic TM running in $O(n)$ time is equivalent to
  a functional finite-state transducer.
\end{proposition}
The Borodin-Trakhtenbrot Gap Theorem \cite{Trakhtenbrot} states that
expanded resources do not always expand the set of computable
functions.  In other words, it is possible that $O(n)$ is
unnecessarily tight time bound for finite-state equivalence.  A less
tight time bound is now expressed with the little-o notation: $t(n)
\in o(f(n))$ means that the upper bound $f(n)$ grows much faster than
the running time $t(n)$ when $n$ tends to infinity:
$\lim_{n\rightarrow \infty} t(n)/f(n) = 0$.

\newcite{Hartmanis:1968:CCO:321450.321464} and \newcite{Trakhtenbrot}
showed independently that the time resource of a finite-state equivalent
deterministic one-tape TM can be expanded from $O(n)$ to $o(n\log n)$
without expanding the characterized languages.  More recently,
\newcite{Tadaki-et-al-2010} showed that the bound $o(n\log n)$ applies
also to nondeterministic one-tape TMs that explore all accepting
computations.
\begin{proposition}[Tadaki et al. 2010]
  A one-tape TM running in $o(n \log n)$ time is equivalent to a
  finite automaton/transducer.
\end{proposition}
A sufficient condition for finite-state equivalence of a TM is
satisfied if the running time of the machine is bounded by a function
$t(n)$ that is in $o(n \log n)$.  For any reasonable function $t(n)$,
this sufficient condition is decidable \cite{Gajser-2015}.  However,
to decide finite-state equivalence of any TM, it would be necessary to
consider all functions $t(n) \in o(n\log n)$.  

We will assume a one-tape TM implementation for finite-fertility
SCGs.  The tape is initialized in such a way that $f$ empty squares
are reserved for latent cohorts at every cohort boundary.

We assume the representation of the grammar rules and the related
application strategy by a functional transducer such as in Figure
\ref{deriv}.  Its optimization via the inward deterministic \emph{bimachine} constructions \cite{YJ2011,Hulden-2011} optimizes the tape moves between derivation steps.

\begin{figure}
  \resizebox{\columnwidth}{!}{\includegraphics{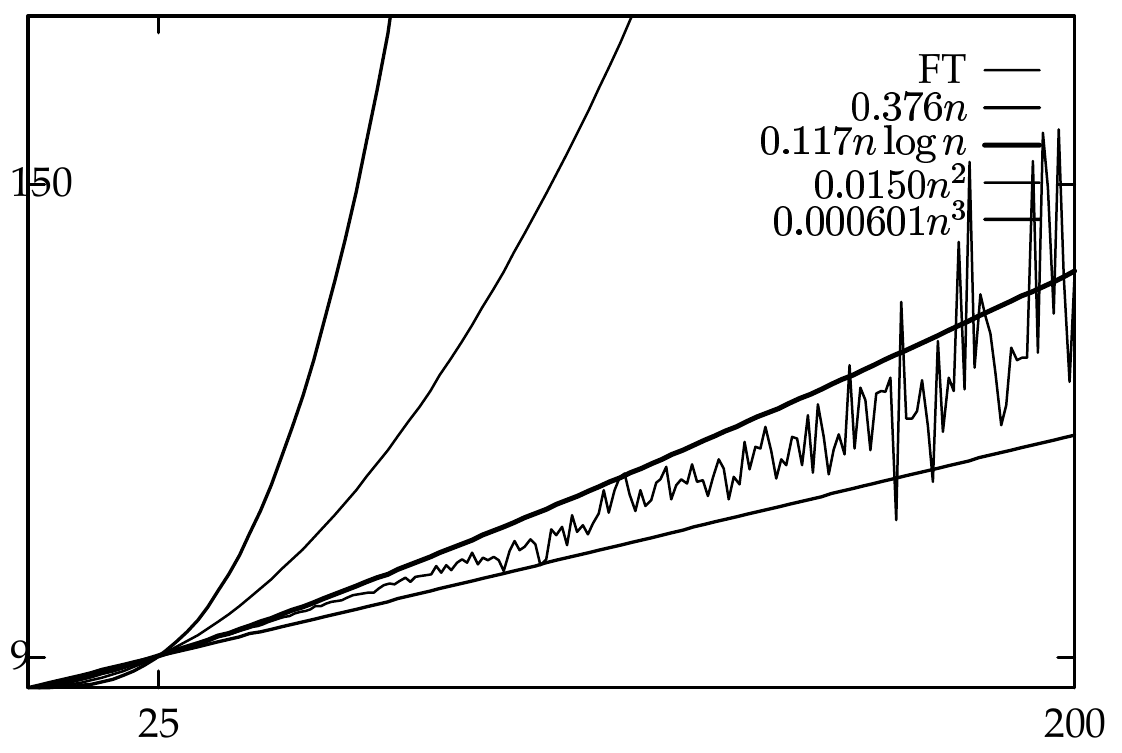}}
\caption{Average running time of CG-2 in Financial Times (according to
  Tapanainen 1999)\label{Tapa} seems to follow the curve $O(n\log n)$}
\end{figure}

The parallel testing of all context conditions involves (i) the
initialization step and (ii) a number of maintenance steps.  The
initialization step computes the validity of all context conditions at
every tape squares in amortised $O(n)$ time.  After this, the total
amortised time needed to maintain the contexts is then bounded by the
total number of moves needed to perform the subsequent rule
applications.
\begin{proposition}
  The time used to maintain the context conditions is dominated by the
  time used to move between target cohorts.
\end{proposition}
NM-SCGs based on a one-tape TM have now a regularity condition:
\begin{proposition}
  An NM-SCG is equivalent to a finite automaton/transducer if its
  one-tape TM implementation runs in $o(n \log n)$ time.
\end{proposition}
This proposition can be compared to an interesting empirical
observation by \newcite{Tapanainen-1999:parintwofra:phd} who reports
experiments with a practical SCG (CG-2) system.  According to the
experiments, the average running time of the system follows closely the
$O(n\log n)$ curve (Fig. \ref{Tapa}).

On the basis of the experiments by Tapanainen, we cannot exclude the
hypothesis that the asymptotic running time is actually in $o(n\log
n)$.  Whether the used grammar is actually equivalent to a
finite-state transducer is not known.

If the given NM-SCG instance would be equivalent to a finite-state
transducer, there would be a possibility to carry out monotonic SCG
parsing in \emph{linear time} and thus improve the parser's efficiency
considerably.  In case that the transducer is extremely large, the
improvement remains solely as a theoretical possibility but the
discovered regularity may still give valuable insight.

\subsection{The $O(n)$ Time Bound}

Hennie's finite-stateness condition \cite{Hennie-1965} for
deterministic one-tape TMs and its generalization to nondeterministic
one-tape TMs \cite{Tadaki-et-al-2010} are insightful and provide a 
method to construct the equivalent
finite-state transducer when the finite-stateness condition is met.

A \emph{Hennie machine} refers to a one-tape TM whose running time is
$O(n)$.  Hennie analysed the expressive power of such machines using
the concept of \emph{crossing sequence}, aka \emph{schema}
\cite{Rabin1963,Trakhtenbrot}.  This concept is a powerful tool in the
analysis of the behaviour of two-way automata and one-tape TMs.

A \emph{crossing sequence} is the sequence of target states $s_1,
s_2,...$ visited by a TM when its pointer crosses the boundary
between a pair of adjacent tape squares.  States $s_1, s_3,...$ are
reached when the pointer moves forward and states $s_2, s_4,...$ are
reached when pointer moves backwards.  Figure \ref{cs} shows how
states are visited during a computation.  The crossing sequence
between the 3rd and the 4th squares is $(s_1,s_2,s_3)=(q_3,q_6,q_9)$.

Every Hennie machine satisfies the property that the length of its
crossing sequences is bounded by an integer $k\in \mathbb{N}$.  
The finiteness of the crossing sequences of a given TM is undecidable
\cite{Prusa} but if a finite upper bound $k$ exists, this constant is
computable \cite{Kobayashi-1985,Tadaki-et-al-2010}.

Finiteness of crossing sequences implies that the TM is equivalent to a finite-state automaton/transducer.  Furthermore, the bound lets us construct this finite-state device.  Unfortunately, the size complexity
of the constructed machine is large in comparison to the original TM:
\begin{proposition}[Pr{\r{u}}{\v{s}}a 2014]
  Each $|Q|$-state, $|\Omega|$-symbol deterministic Hennie
  machine can be simulated by a nondeterministic finite automaton with
  $2^{O(|\Omega| \log |Q|)}$ states.
\end{proposition}

\begin{figure}
\[
{\begin{array}{|c|c|c|c|c|c|c|c|c|ccc}
  \hline
  \texttt{LB} & 
  \texttt{W} & 
  \texttt{O} & 
  \texttt{R} & 
  \texttt{D} & 
  \texttt{I} & 
  \texttt{N} & 
  \texttt{G} & 
  \texttt{RB} &
  \cdots\\
  \hline
  \multicolumn{1}{c}{q_0} &
  \multicolumn{1}{c}{q_1} &
  q_2 &  \multicolumn{1}{c}{\fbox{$q_3$}} &  \multicolumn{1}{c}{q_4} 
  \\
  \multicolumn{1}{c}{} &
  \multicolumn{1}{c}{q_7} &
  \fbox{$q_6$} & \multicolumn{1}{c}{{q_5}} & \multicolumn{1}{c}{\hookleftarrow} 
  \\
  \multicolumn{1}{c}{} &  \multicolumn{1}{c}{\hookrightarrow} &
  q_8 & 
  \multicolumn{1}{c}{\fbox{$q_{9}$}} &
  \multicolumn{1}{c}{{q_{10}}} &
  \multicolumn{1}{c}{{q_{11}}} &
  \multicolumn{1}{c}{\!\!\dots\!\!} 
\end{array}}
\]
\caption{\label{cs}A crossing sequence between squares}
\end{figure}

Testing the finite-stateness of already constructed TMs requires more
effort than to design and construct machines that are immediately known
to be Hennie machines.  We will now mention a few immediate constructions.

\newcite{Prusa}'s construction is based on a
finite weight $w\in \mathbb{N}$ of the tape squares.  Every time when
a square is visited or passed, the weight associated with the square
is reduced.  Once the weight is zero, further visits to the square are blocked.
\begin{proposition}[Pr{\r{u}}{\v{s}}a 2014]
  A weight-reducing one-tape TM is a Hennie machine.
\end{proposition}
Analogously, we can define an NM-SCG whose cohorts has a weight $w$
that is reduced whenever the pointer of the associated TM
implementation visits the corresponding square.  The cohorts of such an NM-SCG have
obviously a finite volume $v\le w$ and can be changed at most $w$
times.
\begin{proposition}
  A finite-fertility NM-SCG implemented by a weight-reducing one-tape
  TM is equivalent to a finite-state transducer.
\end{proposition}
The second way to construct a Hennie-machine based NM-SCG is to set
the maximum distance $m\in \mathbb{N}\cup\{\infty\}$ between adjacent
rule applications.\footnote{This approach was pursued and developed
  further by the current author in an earlier manuscript \cite{YJ2010}
  that is available on request.}  When combined with the linear bound
for rule applications, we obtain the $O(n)$ bound and finite-state
equivalence:
\begin{proposition}
  A finite-fertility NM-SCG runs in $O(m(f+1)vn)$ time and is
  equivalent to a finite-state transducer if $m,f,v \in \mathbb{N}$.
\end{proposition}
The third way is to assume fertility $f\in\mathbb{N}$ and $w=1$.
Since no square can be revisited, this forces the SCG to move
constantly into one direction after all rule applications.  This
special case resembles the rewriting rules in finite-state phonology
whose fundamental theorem
\cite{Johnson-1972:foraspofpho:boo,Kaplan-Kay-1994:regmodofpho:art}
states that if a phonological rule does not reapply to its own output
(but instead moves on), it is regular.

The fourth way to construct a Hennie machine from an SCG is based on
the number of times the context conditions for a cohort has to be
updated.  A monotonic SCG reduces the ambiguity of the sentence at
every rule application.  The reduced ambiguity causes occasional
updates in context conditions of cohorts.  Depending on the
context conditions, such updates at a cohort boundary may have an infinite or finite bound.  Due to functionality and inward determinism of the $\Rightarrow$-transducer, the pointer moves from one cohort to another only if the context conditions 
of the latter have changed as a result of a rule application.
Thus, the number of context updates bound the number of moves:
\begin{proposition}
  If the context conditions can be updated only finitely often at
  every cohort, then the SCG is equivalent to a finite-state
  transducer.
\end{proposition}

\section{Open Problems}

\subsection{Aperiodic Context Conditions}

\newcite{YliJyra-2003:dessynwitsta:inp} showed that the context
conditions used in a realistic Finite-State Intersection Grammar
(FSIG) are not only regular but star-free.  Since context conditions
of SCG rules are strictly weaker than those of FSIG
\cite{Tapanainen-1999:parintwofra:phd}, we have a strong conjecture
that contexts in practical SCG are also star-free.

Star-free languages are definable in the monadic first-order logic of
order, FO$[<]$, a decidable logic that is equivalent to LTL
\cite{Pnueli-1977} and loop-free alternating finite automata (LF-AFA)
\cite{Salomaa-Yu-2000:altfinautand:art}.  The states in an LF-AFA are
totally ordered in such a way that every state is independent from all the preceding states in this order.  This is a major restriction to the structure
and expressive power of alternating finite automata.  

While preserving possible star-freeness has led improvements in fundamental algorithms \cite{YliJyra-Koskenniemi-2004:comconreson:inp},
we have not been able to solve the following open problem:
\begin{openproblem}
  Determine whether the construction of Hennie machines could benefit
  from star-freeness of the context conditions, possibly in combination with other conditions.
\end{openproblem}

\subsection{Full Parsing}

Reductionistic parsing 
\cite{Koskenniemi-1990:finparanddis:inp,Maruyama-1990:bfstrdiswit:inp,Voutilainen-Tapanainen-1993:ambresina:inp,Gross-1997:theconofloc:inc,eisner-smith:2005:IWPT}
is closely related to the consistency enforcing methods used in image
recognition \cite{Huffman,Clowes} and to the satisfiability in logic
\cite{Inari2016}.  All these methods use some idea of 
domains that are then constrained.

\newcite{Karlsson-1990} introduced the term \emph{cohort} for
ambiguity domains or lists of readings associated with tokens.  Lauri
Karttunen has then proposed (p.c., see also Voutilainen
1994)\nocite{Voutilainen-1994:desapargra:boo} that the cohorts
can be treated as strings and processed by finite-state transducers.
This idea has been implemented later by others
\cite{Peltonen-2011,Hulden-2011}.  

Interestingly, the idea of processing ambiguity domains, i.e. cohorts, as strings is actually older than the
SCG tradition.  In the context of formal language theory, it dates
back to \newcite{Greibach} and has been appreciated recently, e.g. by
\newcite{tOkhotin_Alexander13a}.  What is interesting in Greibach's original
use of cohorts is that these cohorts are used to represent parse trees instead
of just morphological ambiguity.  The decomposition of trees and
digraphs into local trees in the lexicon is actually due to the tradition of Categorial Grammar
\cite{ADJ,BH,Lambek-1958}. This suggests an avenue for
future SCG-related research.

\begin{openproblem}
  Develop an SCG grammar that performs full parsing on the basis of
  the structural ambiguity encoded into lexical categories.
\end{openproblem}

\section{Conclusions}

In this paper, the author has laid foundations for the analysis of the
generative power of SCGs.
\begin{itemize}\setlength\itemsep{0em}
\item The parsing is viewed as a derivation that resembles that of
  Generative Phonology.
\item The equivalence between Nonmonotonic SCG and Turing machines is
  established, thus linking Constraint Grammar to Undecidability and
  the Chomsky hierarchy.
\item Finite-fertility SCGs are shown to be context sensitive and
  running in quadratic time.
\item A loose time bound $o(n \log n)$ for finite-state equivalent SCG
  instances (running on a TM) is provided and related to prior
  experiments.
\item Specific conditions for constructing finite-state equivalent
  SCGs are given.
\item Two open problems related to the potential of the star-freeness
  restriction of context conditions and the structural categories in
  the lexicon are presented.
\end{itemize}
The current work has demonstrated that the SCG formalism is not just a
programming language for text linguistics but a formal framework that
lends itself to connections to the
richness of formal language theory and rigorous formal analysis of 
the related parsing complexities, culminating to attempts to reduce grammars into finite transducers.

\section*{Acknowledgements}

The author has received funding as Research Fellow from the Academy of
Finland (dec. No 270354 - A Usable Finite-State Model for Adequate
Syntactic Complexity) and Clare Hall Fellow from the University of
Helsinki (dec. RP 137/2013).  The distance-based restriction
of SCG has been studied by the author \cite{YJ2010} under earlier funding from the first agency (dec. 128536).

\bibliography{sele,cubic,cghennie,cg,fsm,weighted,scattered,min,regex,logic,labeling,bantu}

\begin{thebibliography}{}

\bibitem[\protect\citename{Ajdukiewicz}1935]{ADJ}
Kazimierz Ajdukiewicz.
\newblock 1935.
\newblock Die syntaktische konnexit\"at.
\newblock In Storrs McCall, editor, {\em Polish Logic 1920-1939}, page
  207–231. Oxford University Press, Oxford.
\newblock Translated from Studia Philosophica, 1, 1-27.

\bibitem[\protect\citename{Bar-Hillel}1953]{BH}
Yehoshua Bar-Hillel.
\newblock 1953.
\newblock A quasi-arithmetical notation for syntactic description.
\newblock {\em Language}, 29:47–58.

\bibitem[\protect\citename{{B\"uchi}}1960]{Buchi-1960:weasecariand:art}
J.~R. {B\"uchi}.
\newblock 1960.
\newblock Weak second-order arithmetic and finite automata.
\newblock {\em Zeitschrift f\"ur Mathematische Logik und Grundlagen der
  Mathematik}, 6:66--92.

\bibitem[\protect\citename{{Chanod} and
  {Tapanainen}}1995]{Chanod-Tapanainen-1995:alexintfor:tec}
Jean-Pierre {Chanod} and Pasi {Tapanainen}.
\newblock 1995.
\newblock A lexical interface for finite-state syntax.
\newblock {MLTT} technical report, Rank Xerox Research Centre, Grenoble
  Laboratory, Grenoble, France, February 9.

\bibitem[\protect\citename{Chomsky and Halle}1968]{SPE}
Noam Chomsky and Morris Halle.
\newblock 1968.
\newblock {\em The Sound Pattern of English}.
\newblock Harper \& Row, New York.

\bibitem[\protect\citename{{Chomsky}}1965]{Chomsky-1965:aspofthethe:boo}
Noam {Chomsky}.
\newblock 1965.
\newblock {\em Aspects of the Theory of Syntax}.
\newblock MIT Press, Cambridge, Massachusetts.

\bibitem[\protect\citename{Clowes}1971]{Clowes}
M.~B. Clowes.
\newblock 1971.
\newblock On seeing things.
\newblock {\em Artificiul Intelligence}, 2:79--116.

\bibitem[\protect\citename{Didriksen}2017]{Didriksen-2017}
Tino Didriksen, 2017.
\newblock {\em {Constraint {Grammar} Manual: 3rd version of the {CG} formalism
  variant}}.
\newblock GrammarSoft ApS, Denmark.

\bibitem[\protect\citename{Eisner and Smith}2005]{eisner-smith:2005:IWPT}
Jason Eisner and Noah~A. Smith.
\newblock 2005.
\newblock Parsing with soft and hard constraints on dependency length.
\newblock In {\em Proceedings of the Ninth International Workshop on Parsing
  Technology}, pages 30--41, Vancouver, British Columbia, October. Association
  for Computational Linguistics.

\bibitem[\protect\citename{{Elgot}}1961]{Elgot-1961:decprooffin:art}
Calvin~C. {Elgot}.
\newblock 1961.
\newblock Decision problems of finite automata design and related arithmetics.
\newblock {\em Transactions of the American Mathematical Society},
  98(1):21--51.

\bibitem[\protect\citename{Gajser}2015]{Gajser-2015}
David Gajser.
\newblock 2015.
\newblock {\em Verifying Time Complexity of {T}uring Machines}.
\newblock {Ph.D.} thesis, University of Ljubljana, Department of Mathematics,
  Ljubljana, Slovenia.

\bibitem[\protect\citename{Greibach}1973]{Greibach}
Sheila Greibach.
\newblock 1973.
\newblock The hardest context-free language.
\newblock {\em SIAM Journal on Computing}, 2(4):304--310.

\bibitem[\protect\citename{{Gross}}1997]{Gross-1997:theconofloc:inc}
Maurice {Gross}.
\newblock 1997.
\newblock The construction of local grammars.
\newblock In Emmanuel {Roche} and Yves {Schabes}, editors, {\em Finite-State
  Language Processing}, chapter~11, pages 329--354. A Bradford Book, the MIT
  Press, Cambridge, MA, USA.

\bibitem[\protect\citename{Hartmanis}1968]{Hartmanis:1968:CCO:321450.321464}
Juri Hartmanis.
\newblock 1968.
\newblock Computational complexity of one-tape {T}uring machine computations.
\newblock {\em J. ACM}, 15(2):325--339, April.

\bibitem[\protect\citename{Hennie}1965]{Hennie-1965}
Frederick~C. Hennie.
\newblock 1965.
\newblock One-tape, off-line {T}uring machine computations.
\newblock {\em Information and Control}, 8(6):553--578.

\bibitem[\protect\citename{Huffman}1971]{Huffman}
D.~A. Huffman.
\newblock 1971.
\newblock Impossible objects as nonsense sentences.
\newblock In B.~Meltzer and D.~Michie, editors, {\em Machine Intelligence},
  volume~6, pages 295--323. Edinburgh University Press, Edinburgh, Scotland.

\bibitem[\protect\citename{Hulden}2011]{Hulden-2011}
Mans Hulden.
\newblock 2011.
\newblock Constraint {Grammar} parsing with left and right sequential finite
  transducers.
\newblock In {\em Proceedings of the 9th International Workshop on Finite State
  Methods and Natural Language Processing ({FSMNLP} 2011)}, pages 39--47,
  Blois, France, July. Association for Computational Linguistics.

\bibitem[\protect\citename{Johnson}1972]{Johnson-1972:foraspofpho:boo}
C.~Douglas Johnson.
\newblock 1972.
\newblock {\em Formal Aspects of Phonological Description}.
\newblock Number~3 in Monographs on linguistic analysis. Mouton, The Hague.

\bibitem[\protect\citename{{Kaplan} and
  {Kay}}1994]{Kaplan-Kay-1994:regmodofpho:art}
Ronald~M. {Kaplan} and Martin {Kay}.
\newblock 1994.
\newblock Regular models of phonological rule systems.
\newblock {\em Computational Linguistics}, 20(3):331--378, September.

\bibitem[\protect\citename{Karlsson}1990]{Karlsson-1990}
Fred Karlsson.
\newblock 1990.
\newblock Constraint {G}rammar as a framework for parsing unrestricted text.
\newblock In H.~Karlgren, editor, {\em Proceedings of the 13th International
  Conference of Computational Linguistics}, volume~3, pages 168--173, Helsinki.

\bibitem[\protect\citename{{Karttunen}}1994]{Karttunen-1994:conlextra:inp}
Lauri {Karttunen}.
\newblock 1994.
\newblock Constructing lexical transducers.
\newblock In {\em 15th COLING 1994, Proceedings of the Conference}, volume~1,
  pages 406--411, Kyoto, Japan.

\bibitem[\protect\citename{Kobayashi}1985]{Kobayashi-1985}
K.~Kobayashi.
\newblock 1985.
\newblock On the structure of one-tape nondeterministic {T}uring machine time
  hierarchy.
\newblock {\em Theoretical Computer Science}, 40(2--3):175--193.

\bibitem[\protect\citename{{Koskenniemi}}1990]{Koskenniemi-1990:finparanddis:inp}
Kimmo {Koskenniemi}.
\newblock 1990.
\newblock Finite-state parsing and disambiguation.
\newblock In Hans {Karlgren}, editor, {\em 13th COLING 1990, Proceedings of the
  Conference}, volume~2, pages 229--232, Helsinki, Finland, August.

\bibitem[\protect\citename{Kuroda}1964]{Kuroda-1964}
Sige-Yuki Kuroda.
\newblock 1964.
\newblock Classes of languages and linear-bounded automata.
\newblock {\em Information and Control}, 7(2):207--223.

\bibitem[\protect\citename{{Lager} and
  {Nivre}}2001]{Lager-Nivre-2001:parofspetag:inc}
Torbj\"orn {Lager} and Joakim {Nivre}.
\newblock 2001.
\newblock Part of speech tagging from a logical point of view.
\newblock In P.~{de Groote}, G.~{Morrill}, and C.~{Retor\'e}, editors, {\em
  Logical Aspects of Computational Linguistics}, volume 2099 of {\em Lecture
  Notes in Artificial Intelligence}, pages 212--227. Springer-Verlag.

\bibitem[\protect\citename{Lambek}1958]{Lambek-1958}
Joachim Lambek.
\newblock 1958.
\newblock The mathematics of sentence structure.
\newblock {\em American Mathematical Monthly}, 65:154--170.

\bibitem[\protect\citename{Lindberg and Eineborg}1998]{Lindberg1998}
Nikolaj Lindberg and Martin Eineborg.
\newblock 1998.
\newblock Learning {Constraint Grammar}-style disambiguation rules using
  {Inductive Logic Programming}.
\newblock In {\em 36th ACL 1998, 17th COLING 1998, Proceedings of the
  Conference}, Montr\'eal, Quebec, Canada, August 10-14.

\bibitem[\protect\citename{Listenmaa}2016]{Inari2016}
Inari Listenmaa.
\newblock 2016.
\newblock {\em Analysing Constraint Grammar with {SAT}}.
\newblock Licentiate thesis, Chalmers University of Technology and University
  of Gothenburg, Gothenburg, Sweden.

\bibitem[\protect\citename{{Maruyama}}1990]{Maruyama-1990:bfstrdiswit:inp}
Hiroshi {Maruyama}.
\newblock 1990.
\newblock Structural disambiguation with contraint propagation.
\newblock In {\em 28th ACL 1989, Proceedings of the Conference}, pages 31--38,
  Pittsburgh, Pennsylvania, June 6-9.

\bibitem[\protect\citename{Myhill}1960]{LBA}
John Myhill.
\newblock 1960.
\newblock Linear bounded automata.
\newblock Wadd technical note, Wright Patterson AFB, Wright Air Development
  Division, Ohio, June.

\bibitem[\protect\citename{Okhotin}2013]{tOkhotin_Alexander13a}
Alexander Okhotin.
\newblock 2013.
\newblock Inverse homomorphic characterizations of conjunctive and boolean
  grammars.
\newblock Technical Report 1080, Turku Centre for Computer Science, Turku.

\bibitem[\protect\citename{Peltonen}2011]{Peltonen-2011}
Janne Peltonen.
\newblock 2011.
\newblock Finite state constraint grammar parser.
\newblock In Eckhard Bick, Kristin Hagen, Kaili M\"u\"urisep, and Trond
  Trosterud, editors, {\em Proceedings of the NODALIDA 2011 workshop Constraint
  Grammar Applications, May 11, 2011}, volume~14 of {\em NEALT Proceedings
  Series}, Riga, Latvia.

\bibitem[\protect\citename{Peters and Ritchie}1973]{Peters-Ritchie-1973}
P.~S. Peters and R.~W. Ritchie.
\newblock 1973.
\newblock On the generative power of transformational grammars.
\newblock {\em Information Sciences}, 6:49--83.

\bibitem[\protect\citename{Pnueli}1977]{Pnueli-1977}
Amir Pnueli.
\newblock 1977.
\newblock The temporal logic of programs.
\newblock In {\em Proceedings of the IEEE 18th Annual Symposium on Foundations
  Computer Science}, pages 46--57, New York.

\bibitem[\protect\citename{Pr{\r{u}}{\v{s}}a}2014]{Prusa}
Daniel Pr{\r{u}}{\v{s}}a.
\newblock 2014.
\newblock Weight-reducing {H}ennie machines and their descriptional complexity.
\newblock In Adrian-Horia Dediu, Carlos Mart{\'i}n-Vide, Jos{\'e}-Luis
  Sierra-Rodr{\'i}guez, and Bianca Truthe, editors, {\em Language and Automata
  Theory and Applications: 8th International Conference, LATA 2014, Madrid,
  Spain, March 10-14, 2014. Proceedings}, pages 553--564, Cham. Springer
  International Publishing.

\bibitem[\protect\citename{Rabin}1963]{Rabin1963}
Michael~O. Rabin.
\newblock 1963.
\newblock Real time computation.
\newblock {\em Israel Journal of Mathematics}, 1(4):203--211.

\bibitem[\protect\citename{Ristad}1990]{Ristad:1990:CSG:981823.981853}
Eric~Sven Ristad.
\newblock 1990.
\newblock Computational structure of generative phonology and its relation to
  language comprehension.
\newblock In {\em Proceedings of the 28th Annual Meeting on Association for
  Computational Linguistics}, ACL '90, pages 235--242, Pittsburgh,
  Pennsylvania.

\bibitem[\protect\citename{{Salomaa} and
  {Yu}}2000]{Salomaa-Yu-2000:altfinautand:art}
Kai {Salomaa} and Sheng {Yu}.
\newblock 2000.
\newblock Alternating finite automata and star-free languages.
\newblock {\em Theoretical Computer Science}, 234:167--176.

\bibitem[\protect\citename{Skut \bgroup et al.\egroup
  }2004]{Skut:2004:BCR:1220355.1220384}
Wojciech Skut, Stefan Ulrich, and Kathrine Hammervold.
\newblock 2004.
\newblock A bimachine compiler for ranked tagging rules.
\newblock In {\em Proc. 20th Int'l Conf. on Computational Linguistics}, COLING
  '04, Stroudsburg, PA, USA.

\bibitem[\protect\citename{Tadaki \bgroup et al.\egroup
  }2010]{Tadaki-et-al-2010}
Kohtaro Tadaki, Tomoyuki Yamakami, and Jack C.~H. Lin.
\newblock 2010.
\newblock Theory of one-tape linear-time {T}uring machines.
\newblock {\em Theoretical Computer Science}, 411(1):22--43.

\bibitem[\protect\citename{Tapanainen}1996]{Tapanainen-1996}
Pasi Tapanainen.
\newblock 1996.
\newblock {\em The {Constraint Grammar Parser CG-2}}, volume~27 of {\em
  Publications}.
\newblock Department of General Linguistics, University of Helsinki.

\bibitem[\protect\citename{{Tapanainen}}1999]{Tapanainen-1999:parintwofra:phd}
Pasi {Tapanainen}.
\newblock 1999.
\newblock {\em Parsing in two frameworks: finite-state and functional
  dependency grammar}.
\newblock {Ph.D.} thesis, University of Helsinki, Finland, 1 December.

\bibitem[\protect\citename{Trakhtenbrot}1961]{Trakhtenbrot-1961}
B.~A. Trakhtenbrot.
\newblock 1961.
\newblock Finite automata and logic of monadic predicates.
\newblock {\em Doklady Akademii Nauk SSSR}, 140:326--329.
\newblock In Russian.

\bibitem[\protect\citename{Trakhtenbrot}1964]{Trakhtenbrot}
Boris~A. Trakhtenbrot.
\newblock 1964.
\newblock Turing computations with logarithmic delay (in {R}ussian).
\newblock {\em Albegra i Logica}, pages 33--34.
\newblock English translation in U. of California Computing Center, Tech.
  Report. No. 5, Berkeley, CA, 1966.

\bibitem[\protect\citename{{Voutilainen} and
  {Tapanainen}}1993]{Voutilainen-Tapanainen-1993:ambresina:inp}
Atro {Voutilainen} and Pasi {Tapanainen}.
\newblock 1993.
\newblock Ambiguity resolution in a reductionistic parser.
\newblock In {\em 6th EACL 1993, Proceedings of the Conference}, pages
  394--403, Utrecht, The Netherlands.

\bibitem[\protect\citename{{Voutilainen}}1994]{Voutilainen-1994:desapargra:boo}
Atro {Voutilainen}.
\newblock 1994.
\newblock {\em Designing a Parsing Grammar}.
\newblock Number~22 in Publications of the Department of General Linguistics,
  University of Helsinki. Yliopistopaino, Helsinki.

\bibitem[\protect\citename{{Yli-Jyr\"a} and
  {Koskenniemi}}2004]{YliJyra-Koskenniemi-2004:comconreson:inp}
Anssi~Mikael {Yli-Jyr\"a} and Kimmo {Koskenniemi}.
\newblock 2004.
\newblock Compiling contextual restrictions on strings into finite-state
  automata.
\newblock In Loek Cleophas and Bruce~W. Watson, editors, {\em The Eindhoven
  FASTAR Days, Proceedings}, number 04/40 in Computer Science Reports,
  Eindhoven, The Netherlands, December. Technische Universiteit Eindhoven.
\newblock {\bf[7]}.

\bibitem[\protect\citename{{Yli-Jyr\"a}}2003]{YliJyra-2003:dessynwitsta:inp}
Anssi~Mikael {Yli-Jyr\"a}.
\newblock 2003.
\newblock Describing syntax with star-free regular expressions.
\newblock In {\em 11th EACL 2003, Proceedings of the Conference}, pages
  379--386, Agro Hotel, Budapest, Hungary, April 12--17.

\bibitem[\protect\citename{Yli-Jyr\"a}2011]{YJ2011}
Anssi Yli-Jyr\"a.
\newblock 2011.
\newblock An efficient constraint grammar parser based on inward deterministic
  automata.
\newblock In Eckhard Bick, Kristin Hagen, Kaili M\"u\"urisep, and Trond
  Trosterud, editors, {\em Proceedings of the NODALIDA 2011 workshop Constraint
  Grammar Applications, May 11, 2011}, volume~14 of {\em NEALT Proceedings
  Series}, Riga, Latvia.

\bibitem[\protect\citename{Yli-Jyr\"a}unpublished]{YJ2010}
Anssi Yli-Jyr\"a.
\newblock unpublished.
\newblock Efficient context-sensitive rewriting with inward deterministic
  transducers.
\newblock Manuscript, 11 pages. Archived to EasyChair as a submission to PSC
  2010 (Prague Stringology Conference 2010).

\end{thebibliography}

\end{document}